\documentclass[twocolumn,5p,times]{article}
\usepackage{graphicx}
\usepackage{lscape,graphicx}
\usepackage{hhline}
\usepackage{multicol}
\usepackage{colordvi}
\usepackage{dcolumn}
\usepackage[fleqn]{amsmath}
\usepackage[mathscr]{euscript}
\usepackage[usenames, dvipsnames]{color}
\usepackage[usenames,dvipsnames,svgnames,table]{xcolor}
\usepackage[linktocpage]{hyperref}
\usepackage[a4paper,bindingoffset=0.2in,left=0.5in,right=0.5in,top=1.0in,bottom=1.0in,footskip=0.25in]{geometry}
\usepackage{caption}
\usepackage{subcaption}
\newcommand*{\doi}[1]{\href{http://dx.doi.org/#1}{doi: #1}}
\usepackage[numbers]{natbib}
\makeatletter
\renewcommand{\section}{\@startsection{section}{1}{\z@}{-3.5ex \@plus -1ex \@minus -.2ex}{1.3ex \@plus.2ex}{\normalfont\small\bfseries\boldmath}}
\makeatother

\makeatletter
\renewcommand{\subsection}{\@startsection{subsection}{2}{\z@}{-3.5ex \@plus -1ex \@minus -.2ex}{1.3ex \@plus.2ex}{\normalfont\small\bfseries\boldmath}}
\makeatother

\makeatletter
\renewcommand{\subsubsection}{\@startsection{subsubsection}{3}{\z@}{-3.5ex \@plus -1ex \@minus -.2ex}{1.3ex \@plus.2ex}{\normalfont\small\bfseries\boldmath}}
\makeatother

\makeatletter
\everymath{\if@display\else\thickmuskip=0mu plus 0mu\fi}
\makeatother

\title{\large \bf  {\tt ee$\in$MC}: Low Energy Mesons and the Residual QCD Potential }
\date{}

\author{\normalsize Ian M. Nugent$^{*}$ \\ \normalsize Victoria, B.C., Canada}

\begin{document}
\twocolumn[
  \begin{@twocolumnfalse}
    \maketitle
\begin{abstract}
The Flux-Tube Breaking Model in {\it ee$\in$MC} is expanded to include the residual QCD potential between the Final-State mesons, within the non-relativistic limit. These residual QCD potentials have 
been predicted in the context of the Flux-Tube Breaking Models to generate meson-meson molecular states for the $f_{0}(500)$, $f_{0}(980)$, $a_{0}(980)$, through the colour hyper-fine spin-spin 
interaction. These residual potentials are also found to have an important impact on the $S_{1}$ decay of the $a_{1}$ and $K_{1}$ axial-vector mesons due to the colour hyper-fine spin-spin interaction.  
It is found that in the low mass regions, the $\rho(770)$ and $K^{*}(892)$ are sensitive to the linear-confining potential and colour-Coulomb potential suggesting that with the high statistics
at the B-Factories, it may be possible to probe the linear-confining potential and colour-Coulomb potential through a model dependent description of the resonance shape or by exploiting multiple production 
process.
\\ \\
Keywords: Electron-Positron Collider, Tau Lepton, Monte-Carlo Simulation \\ \\
\end{abstract}
\end{@twocolumnfalse}
]

\renewcommand{\thefootnote}{\fnsymbol{footnote}}
\footnotetext[1]{Corresponding Author \\ \indent   \ \ {\it Email:} inugent.physics@outlook.com}
\renewcommand{\thefootnote}{\arabic{footnote}}

\section{Introduction }

Inter-meson interactions through the residual QCD potential have been proposed in the Flux-Tube-Breaking models as an explanation for the low energy $f_{0}(500)$, $f_{0}(980)$  $a_{0}(980)$, and
 $K_{0}^{*}(700)$ scalar states \cite{Weinstein:1982gc,Weinstein:1983gd,Barnes:1986uu,Weinstein:1990gu,Isgur:1989js}. Within this picture, the residual QCD potential in the Final-State 
forms inter-meson ``molecular'' states \cite{Weinstein:1982gc,Weinstein:1983gd,Barnes:1986uu}. This interpretation is supported by evidence from $\psi$ and $\eta(1440)$ decays in $\pi\pi-KK$ 
scattering and scalar $\gamma\gamma$ coupling couplings \cite{Barnes:1986uu,Barnes:1985cy,ODonnell:1985jss,Weinstein:1985ir}.  The production of the scalar mesons through this mechanism does
 not depend on the chiral symmetry and is therefore an alternative hypothesis to models in which the low mass scalar mesons are directly related to the origin of the quark-composite mass 
through chiral symmetry breaking 
\cite{PDG2020,PhysRevD.59.074001,PEL_EZ_2004,Dabado_1997,Oller_2003,Tornqvist_1999,Napsuciale_2004,Napsuciale_2004b,Ishida_1999,Sadron_1999,Fariborz_2014,Black_1999,GellMann:1960,Skyrme:1962vq,Skyrme:1962vh,Speight:2018,Nambu:1961}.  This provides a complementary alternative hypothesis for the low mass scalars to \cite{Nugent:2022ayu}.
  These wave-function amplitude distortions from Final-State residual QCD meson interactions are not only limited to
 the S-wave channel, but are predicted to play an essential role in low energy ``meson-evolution'' \cite{Isgur:1989js}.
Within the context of the Flux-Tube Breaking Model \cite{Kokoski:1985is,Isgur:1983wj,Isgur:1984bm,Isgur:1988vm}, supplemented with phenomenological models of the residual QCD potential, we extend the 
models to the majority of decays of light quark mesons within the Flux-Tube Breaking Model presented in \cite{Nugent:2022ayu}.   

\section{Low Energy QCD Potential Models \label{sec:ResidualQCD}}

The residual QCD interaction between the outgoing Final-State mesons modifies the cross-section (decay-rate) through both corrections to the meson propagators and wave-function 
amplitude distortion \cite{Weinstein:1982gc,Weinstein:1983gd,Barnes:1986uu}. At low energies the modification to the hadronic propagators are expected to be negligible \cite{Barnes:1986uu}, and will 
therefore be neglected in this work.  This leaves the wave-function amplitude distortions caused by the Final-State interaction between the mesons,

\begin{equation}
\resizebox{0.35\textwidth}{!}{$
S_{h_{1}h_{2}}^{2}=\frac{\sigma\left(h_{0}\to h_{1}h_{2} | V(r)\right)}{\sigma\left(h_{0}\to h_{1}h_{2} \right)}\approx\frac{\left| \psi\left(r=0 | V(r)\right) \right|^{2}}{\left| \psi\left(r=0 \right) \right|^{2}}.
$}
\label{eq:Sab2}
\end{equation}

\noindent It follows from equation \ref{eq:Sab2}, that in the Flux-Tube Breaking Model,
the width of the meson resonances is also dependant on $S_{h_{1}h_{2}}^{2}$, and therefore must be included in the computation of the mass dependent width $\Gamma(s)$, and the effective mass of the
resonances, $\hat{m}^{2}$ and $\bar{m}^{2}$. These amplitude distortions to the wave-function from the effective QCD potential on the outgoing wave-function of the Final-State hadrons can be approximated by 
the non-relativistic relative Hamiltonian for plane waves being scattered by a potential in the non-relativistic static limit \cite{Barnes:1986uu}. Although, the validity of this approximation can be
questioned in the relativistic regime \cite{Barnes:1986uu}, the predominate effect is in the low energy non-relativistic region. In the relativistic regime, 
$S_{h_{1}h_{2}}$ converges to unity. Within this context, the radial component of the wave-function, $\psi(r,\theta,\phi)=R(r)Y_{m}^{l}(\theta,\phi)$, is determined 
numerically \cite{Abramowitz:1964,WehShen:2016} from the Schr\"odinger equations in spherical coordinates,

\begin{equation}
\resizebox{0.4\textwidth}{!}{$
\begin{array}{ll}
R^{\prime\prime}(r)=-\frac{2}{r} R^{\prime}(r)+2\mu\left(\frac{l(l+1)}{2\mu r^{2}}+V(r)-E_{h_{1}h_{2}} \right)R(r)
\end{array}
$}
\label{eq:RadialSchrodinger}
\end{equation}

\noindent where l is the angular momentum and $\mu$ is the reduced mass of the system. The initial conditions correspond to the Spherical Bessel Functions $J_{0}(x)$, $J_{1}(x)$ and $J_{2}(x)$ for 
the S-wave, P-wave and D-wave respectively\cite{Liboff:1997} at the origin of production for the Final-State particles ($r=0/x=0$). From the Flux-Tube Breaking Model, the largest contribution to the
 potential energy for the light quark mesons produced in a S-wave state is due to the colour hyperfine spin-spin interaction \cite{Barnes:1999hs}.  This is described in the Flux-Tube Breaking Model   
by the expectation value of the Hamiltonian.  For a single meson this may be written as: 
\begin{equation}
< \psi_{nS}| H_{hyp} | \psi_{nS} > = \frac{32\hat{S}_{i}\cdot \hat{S}_{j}\pi\alpha_{s}}{9m_{i}m_{j}}|\psi_{ns}(0)|^{2}
\label{eq:hypfine}
\end{equation}
\noindent \cite{Kokoski:1985is,Barnes:1999hs} where $\alpha_{s}=0.6$\cite{Godfrey:1985xj} is the frozen coupling constant, $\hat{S}_{i}$ and $\hat{S}_{j}$ are the spin operators 
for the $i$th and $j$th valence quark from the Initial-State meson. $m_{i}$ and $m_{j}$ correspond to the mass of the $i$th and $j$th valence quark in the non-relativistic static potential, 
$m_{u/d}=0.22GeV/c^{2}$ and $m_{s}=0.419Gev/c^{2}$\cite{Godfrey:1985xj}. For two mesons $\beta_{B}=\beta_{C}=\beta$ in a ground state emitted in a $S_{1}$ configuration, this reduces  

\begin{equation}
 \resizebox{0.3\textwidth}{!}{$ 
V_{hyp}(r ) = \frac{32  \alpha_{s} \hat{S}_{i}\cdot \hat{S}_{j} \beta^{3}}{ 9 (2\pi)^{1/2}m_{i}m_{j} } 
e^{-\frac{(\beta r)^2}{2}}
$}
\label{eq:hypfineS1S1}
\end{equation}

\noindent by means of convolution of the two Gaussians \cite{Arfken}.  For the P and D-wave production, the expectation value of the colour hyperfine spin-spin interaction is not expected to
 contribute based-on the orthogonality of states and symmetry \cite{Scora:1995ty}. For the $K_{1}$ system, the expectation value of the spin-spin operator, $\hat{S}_{i}\cdot\hat{S}_{j}$, 
is not the same between the singlet and triplet states, and therefore must  be incorporated into the mixing of the singlet and triplet states \cite{Nugent:2022ayu,Suzuki:1993,Blundell:1995au,Asner:2000nx}.
 Thus the potential for the 
colour hyper-fine spin-spin interaction may not only be attractive but also repulsive in the $S_{1}$ decays of the $K_{1}(1270)$ and $K_{1}(1400)$ mesons depending on the given  $s$. For the case were
$\beta_{B}\neq\beta_{C}$, the colour hyper-fine spin-spin potential may be written as:  

\begin{equation}
 \resizebox{0.35\textwidth}{!}{$                                                    
V_{hyp}(r ) = \frac{32  \alpha_{s} \hat{S}_{i}\cdot \hat{S}_{j} (\beta_{B}^{3}\beta_{C}+\beta_{B}\beta_{C}^{3})}{ 18 (\pi)^{1/2}\left(\beta_{B}^{2}+\beta_{C}^{2}\right)^{1/2}m_{i}m_{j} }
e^{-\frac{(\beta_{B}\beta_{C} r)^2}{\left(\beta_{B}^{2}+\beta_{C}^{2}\right)}} 
$}
\label{eq:hypfineS1S1}
\end{equation}
\noindent where symmetrization between the wave-functions has been applied for the respective normalizations\footnote{In the results presented here $\beta_{B}=\beta_{C}=0.4$, only $\beta_{A}$ 
in the axial-vector mesons deviates from $\beta=0.4$ in this work with a value of $\beta=0.35$.}.

The remaining potentials in the Flux-Tube Breaking Model, are due to the linear-confining potential,

\begin{equation}
\resizebox{0.4\textwidth}{!}{$
V_{L}^{I=0}=-\frac{C_{I}^{2}b}{3\beta}\left(\beta r +2\sqrt{\frac{2}{\pi}}-\left(\beta r+ \frac{2}{\beta r}\right)Erf\left(\frac{\beta r}{2}\right)  \right) e^{-\frac{(\beta r)^2}{2}} -\frac{2}{\sqrt{\pi}} e^{-\frac{3(\beta r)^2}{4}}
$}
\label{eq:linconf}
\end{equation}

\noindent \cite{Barnes:1999hs} and colour-Coulomb potential, 

\begin{equation}
 \resizebox{0.4\textwidth}{!}{$ 
V_{C}^{I=0}(r)=\frac{4C_{I}^{2}\alpha_{s}}{9r}\left(1+\sqrt{\frac{2}{\pi}}\beta r -4Erf\left(\frac{\beta r}{2}\right)\right)e^{-\frac{(\beta r)^2}{2}}
$}
\label{eq:coulumb}
\end{equation}

\noindent \cite{Barnes:1999hs}\footnote{We note that the $f_{0}(1370)$ and $K_{0}^{*}(1430)$ are ${}^{3}P_{0}$ states and that the wave-function overlap is 
approximated as resulting from S wave particles.}. From the symmetrization of the transition matrix it follows that $V_{C}^{I=0}= -V_{C}^{I=1}$ and $V_{L}^{I=0}= -V_{L}^{I=1}$ \cite{Barnes:1999hs}\footnote{We follow the $I=1$ and $I=0$ sign convention from \cite{Barnes:1999hs}.}.

\begin{figure*}[tbp]
\begin{center}
  \resizebox{500pt}{120pt}{
    \includegraphics{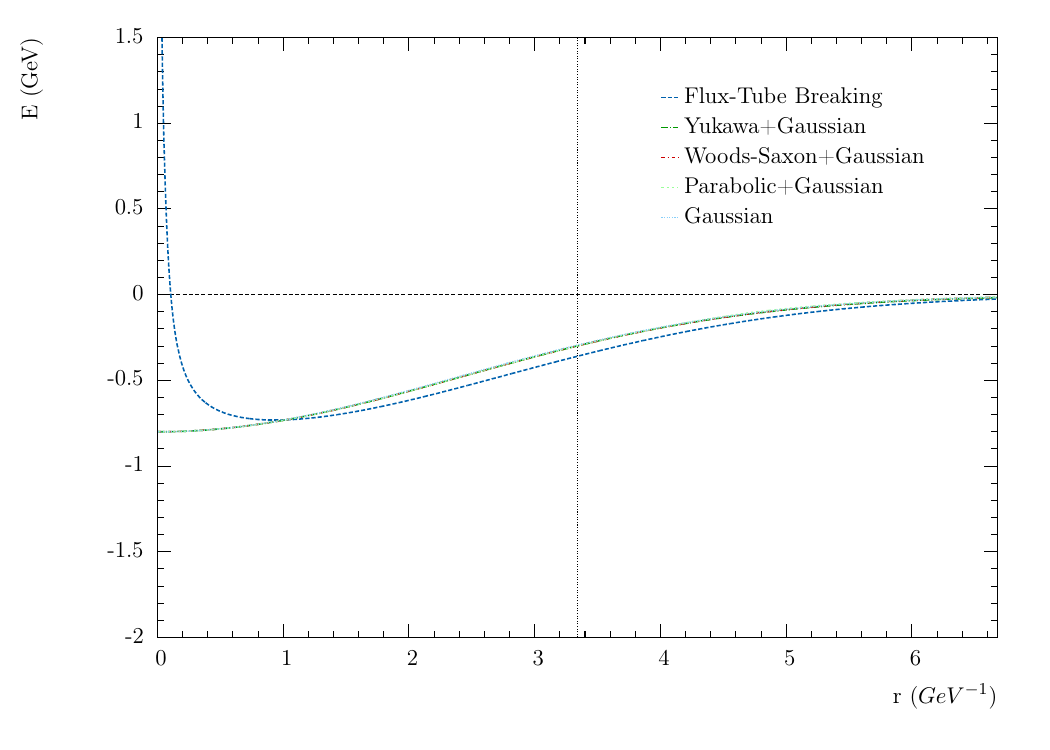}
    \includegraphics{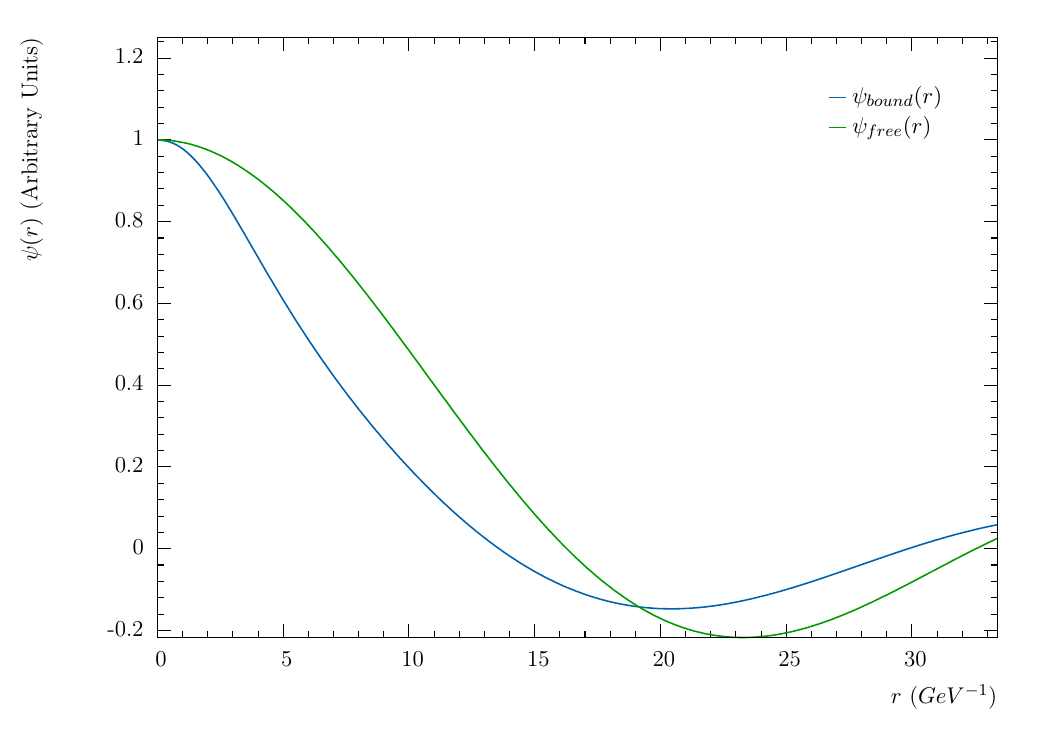}
    \includegraphics{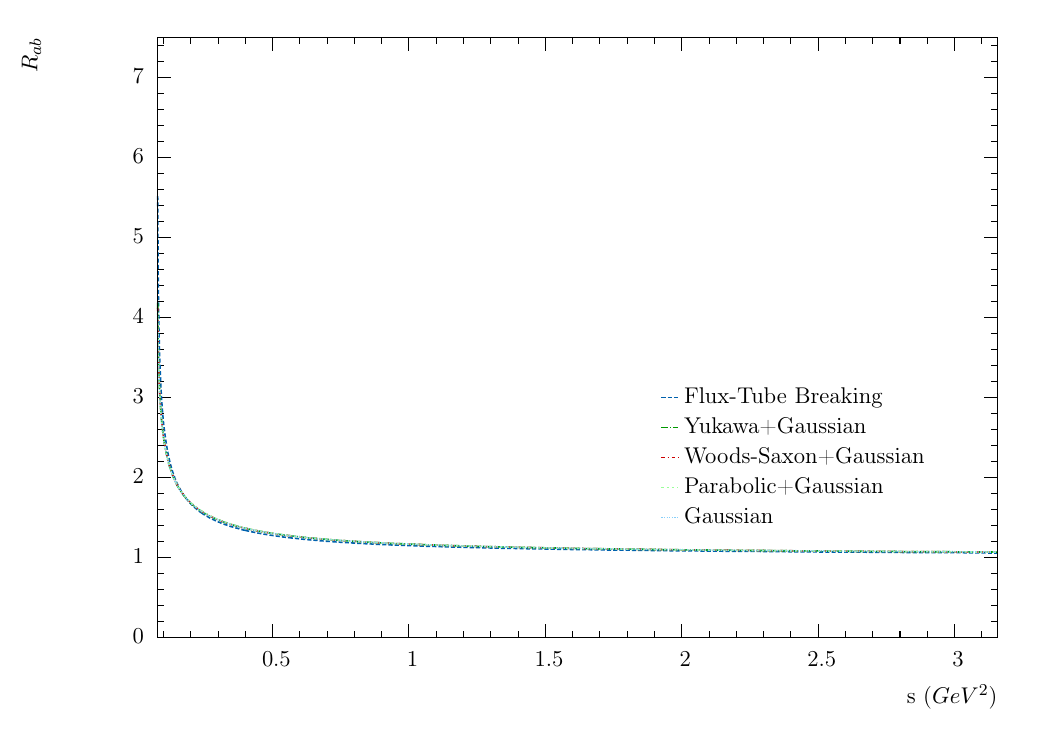}
  }
  \resizebox{500pt}{120pt}{
    \includegraphics{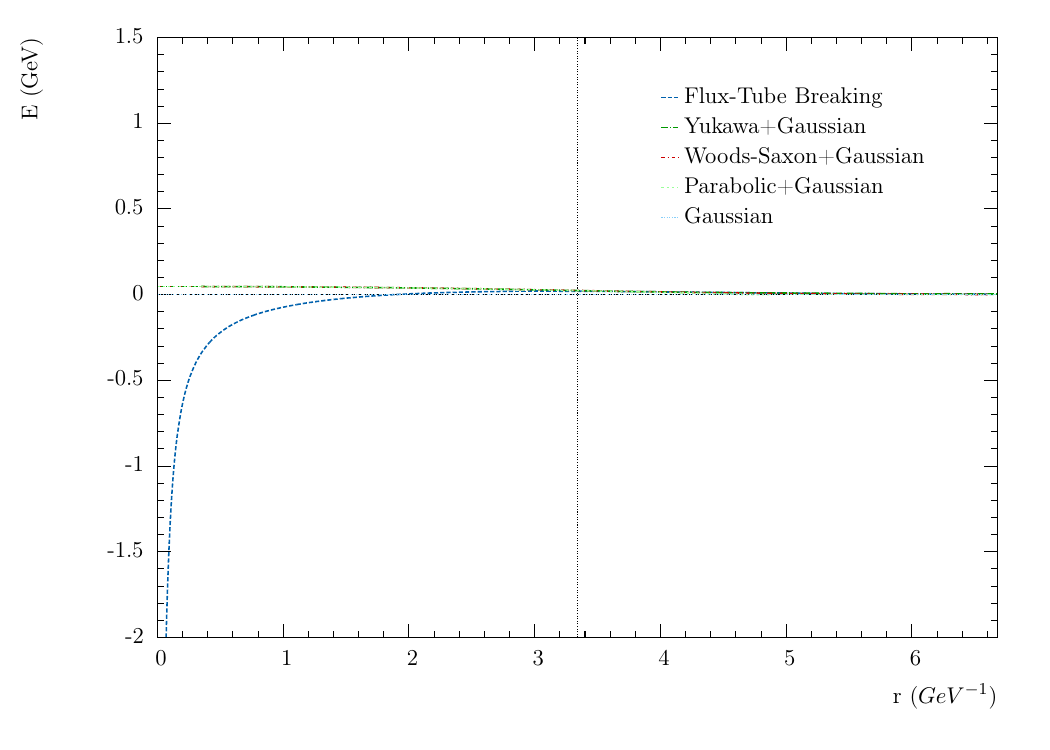}
    \includegraphics{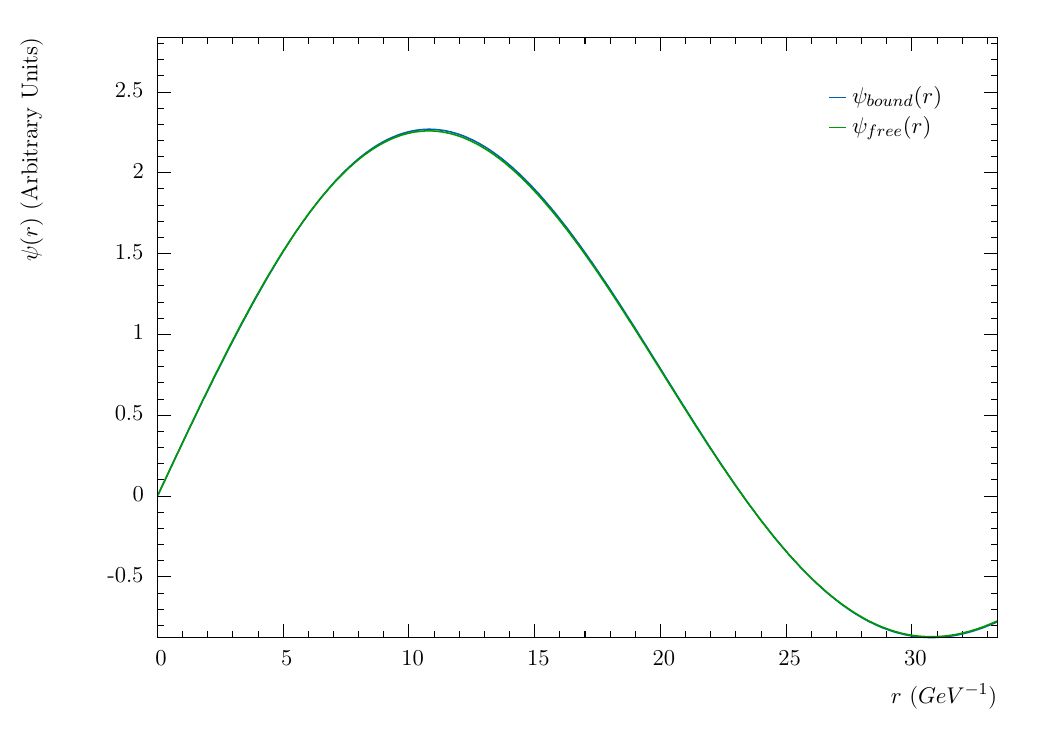}
    \includegraphics{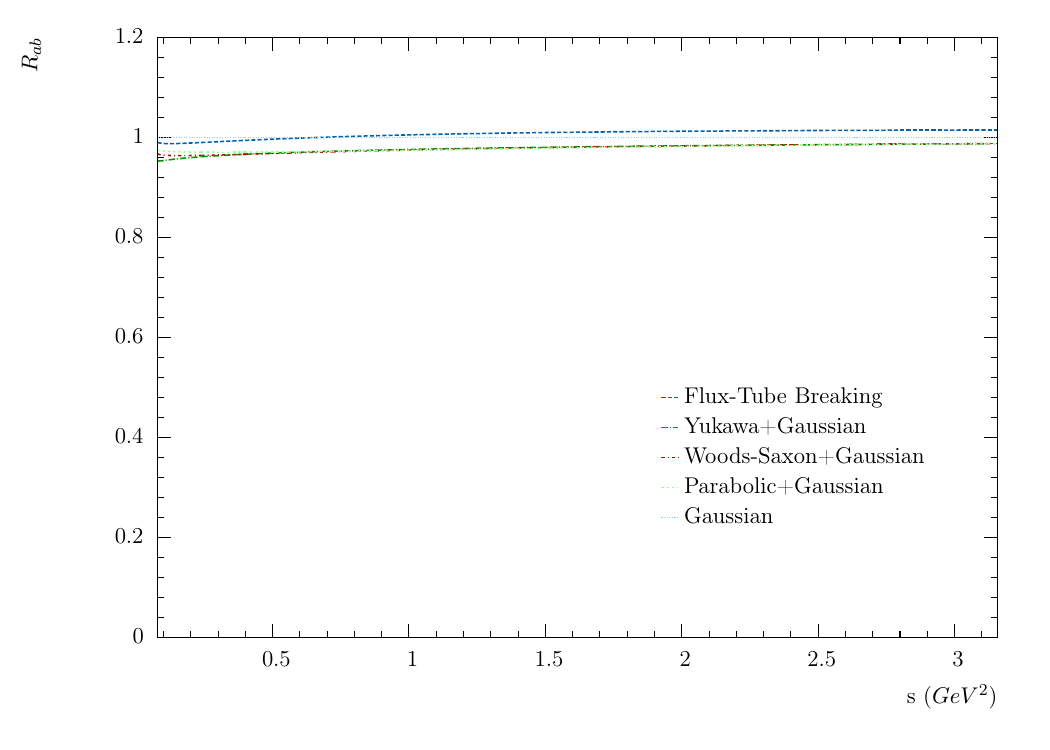}
  }
  \resizebox{500pt}{120pt}{
    \includegraphics{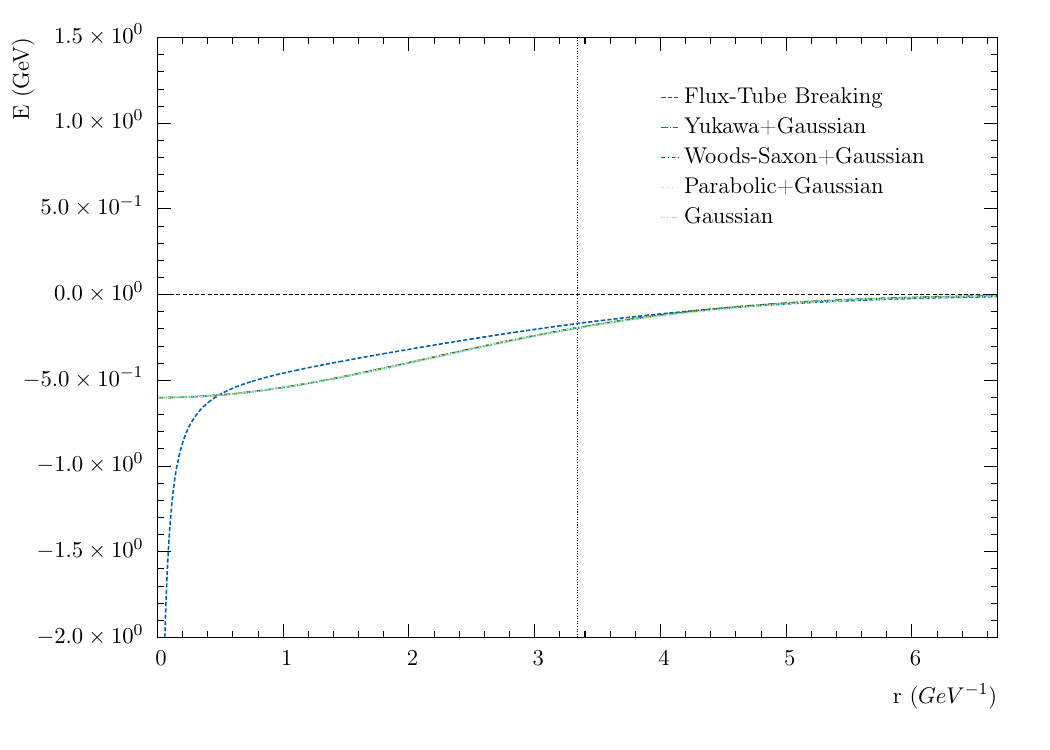}
    \includegraphics{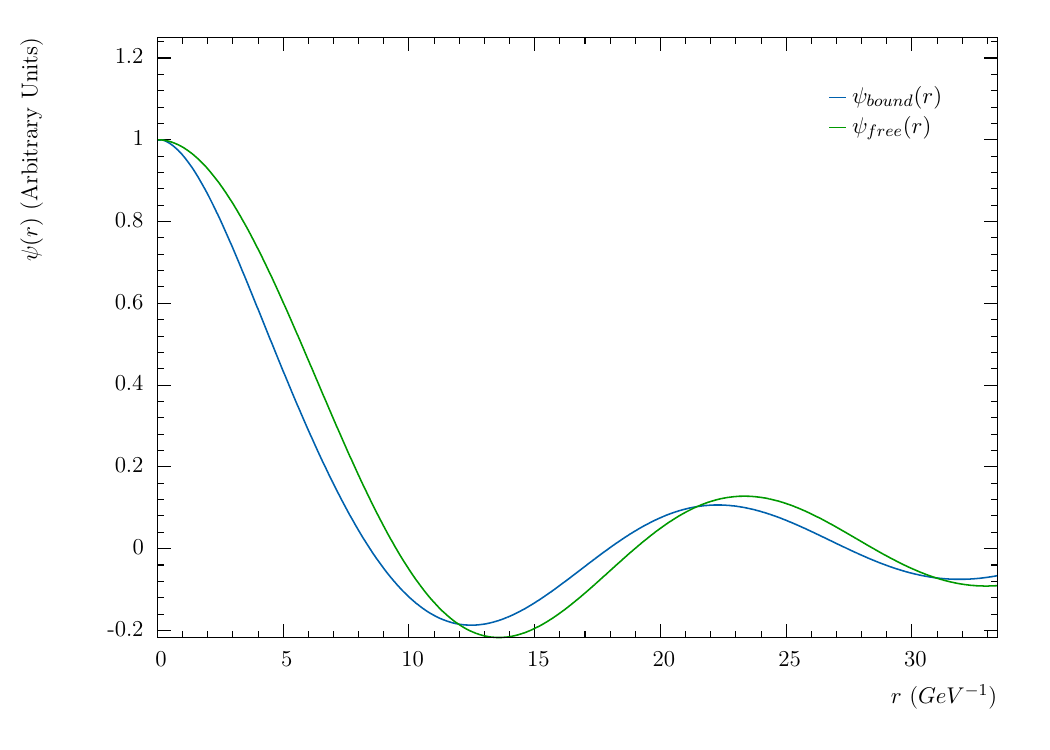}
    \includegraphics{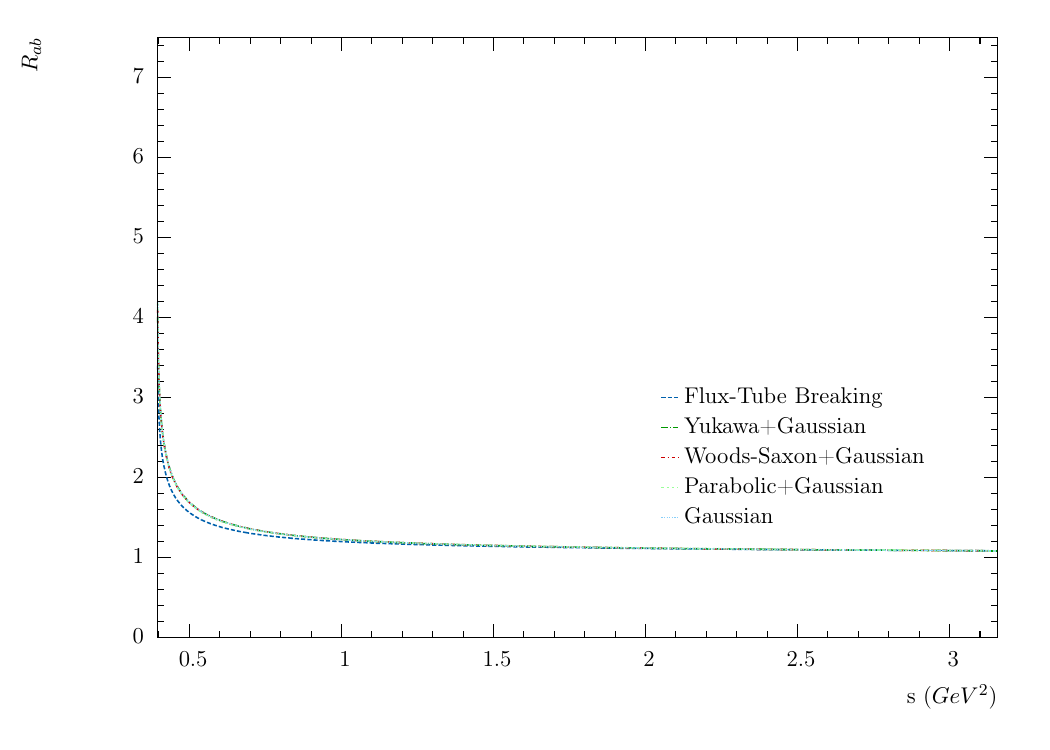}
  }
 \resizebox{500pt}{120pt}{
    \includegraphics{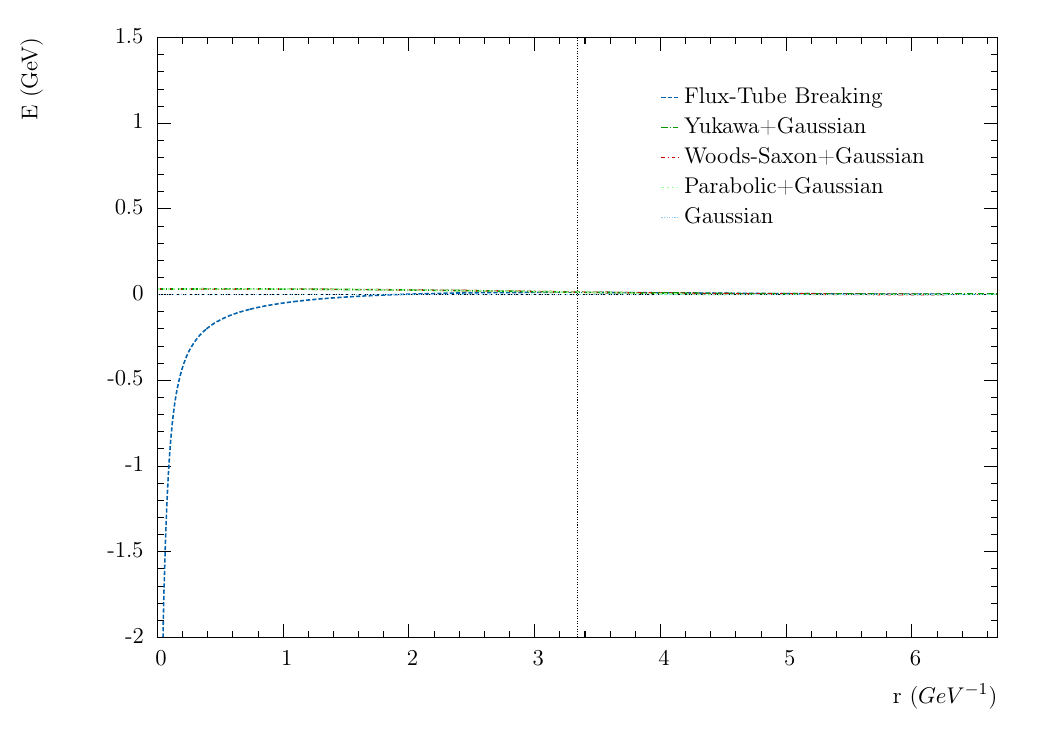}
    \includegraphics{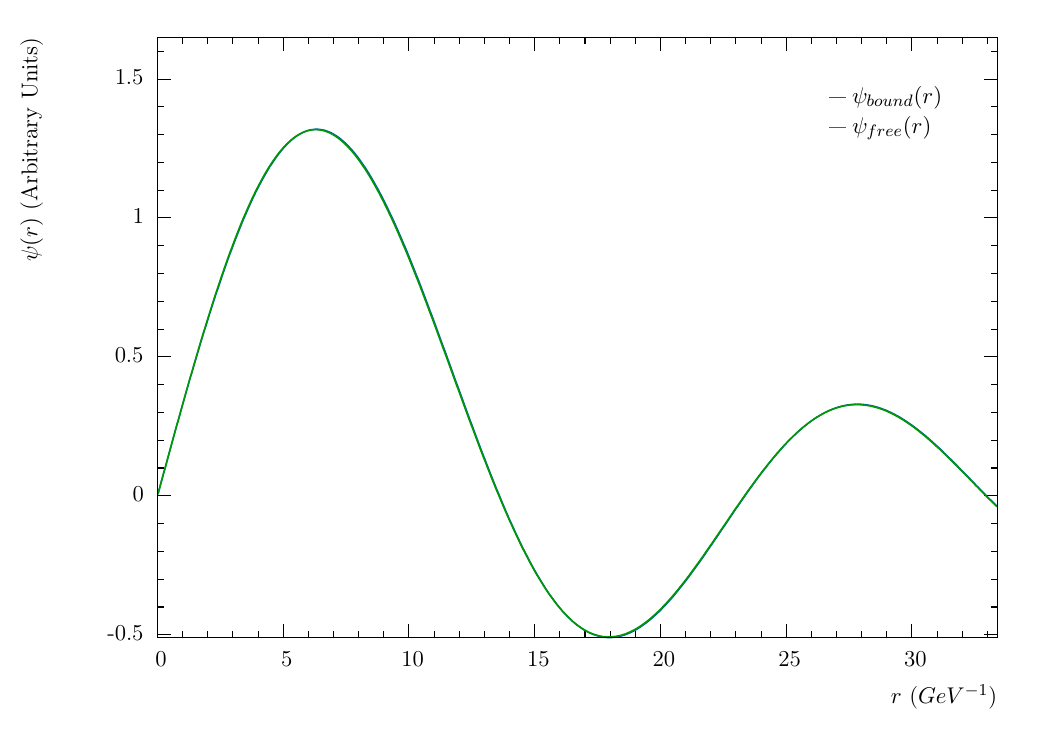}
    \includegraphics{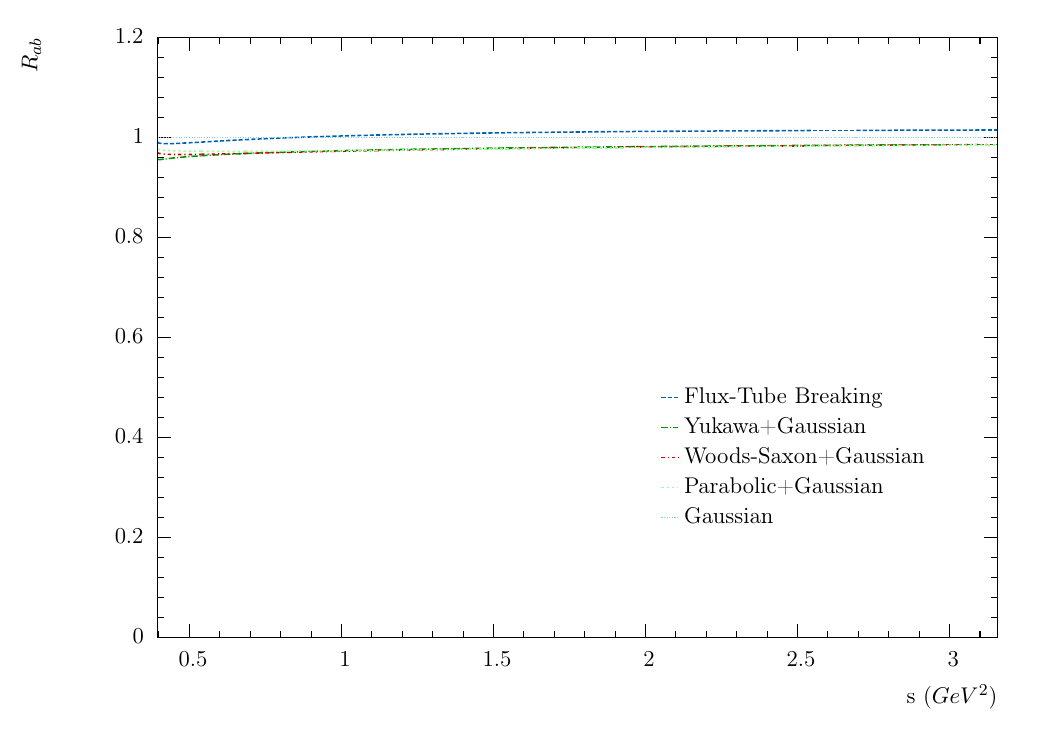}
  }
\end{center}
  \caption{The effective potentials including angular momentum (left), wave-function solutions for the Flux-Tube Breaking Model with  $s=1.25\times s_{thres}$ (middle) and the wave-function distortion factor 
$S_{h_{1}h_{2}}^{2}$ (right) for the effective potential 
models presented in Table \ref{table:Eq} for the $\pi^{+}\pi^{-}$ [$l=0$] (top), [$l=1$] (upper-middle), $K^{-}\pi^{0}$ [$l=0$] (lower-middle), [$l=1$] (bottom) Final-States. 
The angular momentum potential $l(l+1)/(2\mu r^{2})$ is not included in the figures, but has been included in the calculation.  \label{fig:Enhancement}   }
\end{figure*}

From Figure \ref{fig:Enhancement}, it can be seen, that there is a potential for bound states to be formed for the S-wave production, which is qualitatively consistent with \cite{Barnes:1986uu}.
Na\"ively,  one would expect that the bound states would be below the $h_{1}h_{2}$ production 
threshold and would decay 
either electromagnetically or weakly, while other models predict higher mass pseudo-stable bound meson-meson molecular 
states \cite{Sazdjian:2022kaf}\footnote{If the stable bound meson-meson molecular states have a sufficiently long life-time, 
they could be a potential background for dark-matter searches.}.  The bound states are not investigated in this paper, but instead, we will focus on the
scattering of the hadronic plane waves in the continuum which cause the Final-State interaction between the hadrons.

In addition to the Flux-Tube Breaking potentials, we include phenomenological models to investigate the theoretical sensitivity to the core-meson potential and residual external potential. The 
colour hyper-fine spin-spin interaction
potential associated with this Final-State interaction between the mesons is modeled as a sum of the short-range residual QCD potential for each meson. Based on imperial evidence \cite{Barnes:1986uu}, 
and the latter Flux-Tube Breaking Model description, it
is modelled using a Gaussian for the S-wave production channels which is normalized to the values extracted in \cite{Barnes:1986uu}.
For consistency with the Flux-Tube Breaking potentials \cite{Barnes:1986uu}, $r_{0}=a_{0}/\sqrt{2}$ with $a_{0}$ representing the charge radius of the meson.
This is combined with an effective potential for the colour-Coulomb potential  outside of the mesons and the core potential inside the mesons based on the shell model \cite{walecka}. In these models, the 
phenomenological functional form approximates the colour-Coulomb potential for a given colour charge density distributed and is therefore finite\footnote{The colour charge 
density goes to 0 within the meson surface layer which has a mean value of the charge radius\cite{walecka}.}. These models include a parabolic distribution based on the Shell Model \cite{walecka} with
 no external residual QCD potential (Parabolic), a parabolic core distribution with a Yukawa residual QCD potential (Parabolic-Yukawa) and the Woods-Saxon potential for the Shell Model 
\cite{walecka,Woods:1954zz}. The normalization is obtained from the linear-confining and colour-Coulomb terms in the Flux-Tube Breaking Model at $r=\beta$($r=a_{0}$). 
The definition for each of these potentials and the corresponding parameters can be found in Tables \ref{table:Eq} and \ref{table:Para}. 
This gives effective potentials of $\mathcal{O}(10-100MeV)$ which are consistent with nuclear potentials in the Shell Model \cite{walecka}. 

\section{Modification to the Flux-Tube Breaking Model}

The Flux-Tube Breaking Model \cite{Kokoski:1985is,Isgur:1983wj,Isgur:1984bm,Isgur:1988vm,Godfrey:1985xj} implemented in  {\tt ee$\in$MC} \cite{Nugent:2022ayu} is extended to include an enhanced 
 vector decay-width calculation which incorporates higher threshold decay processes while the $f_{0}(1370)$ decay processes are extended to include the $\gamma\gamma$ channel.
This is achieved by replacing the formalism for the  $f_{VPP}$ form-factor from \cite[Eq. A.16]{Isgur:1988vm}, with the formalism in \cite{Kokoski:1985is}\footnote{This includes applying the 
corresponding corrections throughout the program.}. 
 This is particularly important for the $\rho(770)\to KK$ contribution which has an amplitude/A of $\frac{\sqrt{3}}{\sqrt{2}}P_{1}$. 
The non-strange low energy scalar sector in the Flux-Tube Breaking Model \cite{Nugent:2022ayu}, 
in particular, the $f_{0}(500)$ and $f_{0}(980)$, is described in terms of a threshold effect 
from $\pi^{0}\pi^{0}$, $\pi^{+}\pi^{-}$, $K^{+}K^{-}$  and $K^{0}\bar{K}^{0}$ for the $f_{0}(1370)$ meson resulting from the dispersion relation \cite{Nugent:2022ayu}. These threshold effects are sensitive to 
lower mass threshold decay processes. As such, the threshold effects from the $\pi^{0}\pi^{0}$ and $\pi^{+}\pi^{-}$ are expected to be extremely sensitive to the $\gamma\gamma$ channel which has thus 
far not been included.  Therefore the $f_{0}(1370)$ model is extended to incorporate 
the $f_{0}(1370)\to\gamma\gamma$ contribution determined within the Bethe-Salpeter bound state formalism \cite{Bergstrom:1982qv} where the scalar amplitude $A({}_{1}^{3}P_{0}=0^{++})$ is approximated by 
means of the Flux-Tube-Breaking Model \cite{Godfrey:1985xj}. The decay width may be written as:

\begin{equation}
\resizebox{0.35\textwidth}{!}{$
\begin{array}{ll}
\Gamma(s)&= \frac{S|M|^{2}}{16\pi M_{f_{0}(1370)}} \\
&= \frac{|F_{S}(s) \sum_{i}\sum_{j} \epsilon^{\alpha\beta\gamma\delta}\epsilon_{(i)\alpha}^{*}q_{\beta}\epsilon_{(j)\gamma}^{\prime *}q_{\delta}^{\prime}|^{2}}{16\pi M_{f_{0}(1370)}} \\
\end{array}
$}
\label{eq:HadvacBW}
\end{equation}

\noindent where

\begin{equation}
\resizebox{0.27\textwidth}{!}{$
F_{S}(s)=\frac{\sqrt{24\pi}\alpha_{QED}(s) e_{Q}^{2} A({}_{1}^{3}P_{0}) }{M_{f_{0}(1370)}^{3/2} m_{Q}^{2}}.
$}
\label{eq:HadvacBW}
\end{equation}

\begin{figure*}[tbp]
\begin{center}
  \resizebox{500pt}{178pt}{
    \includegraphics{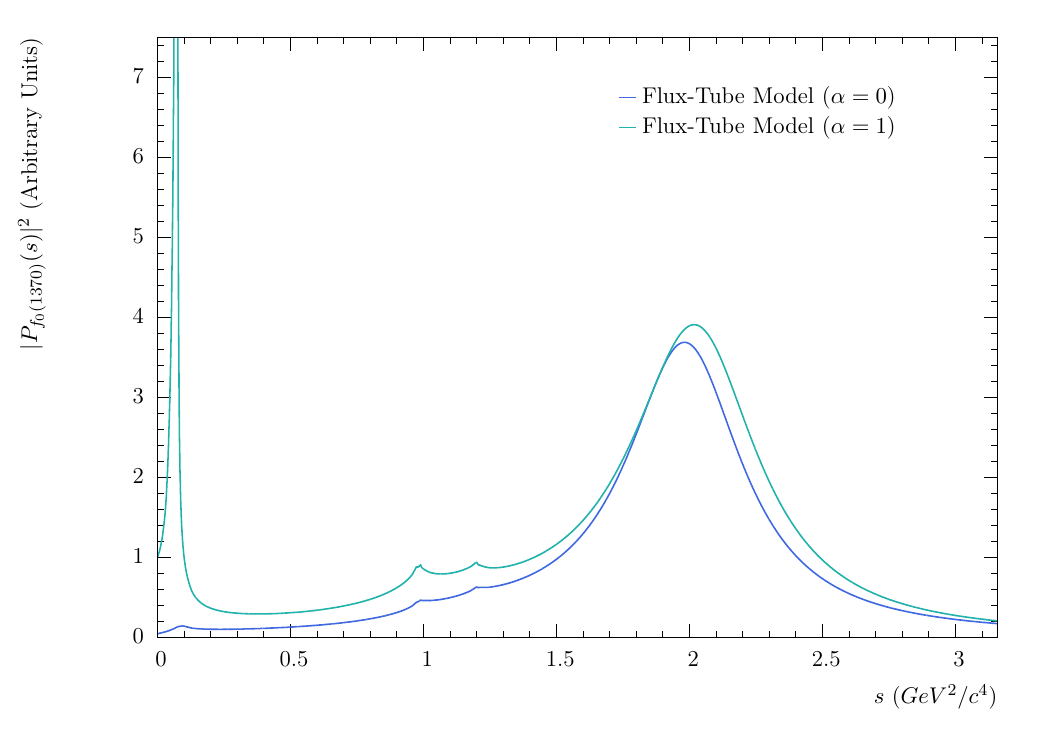}
    \includegraphics{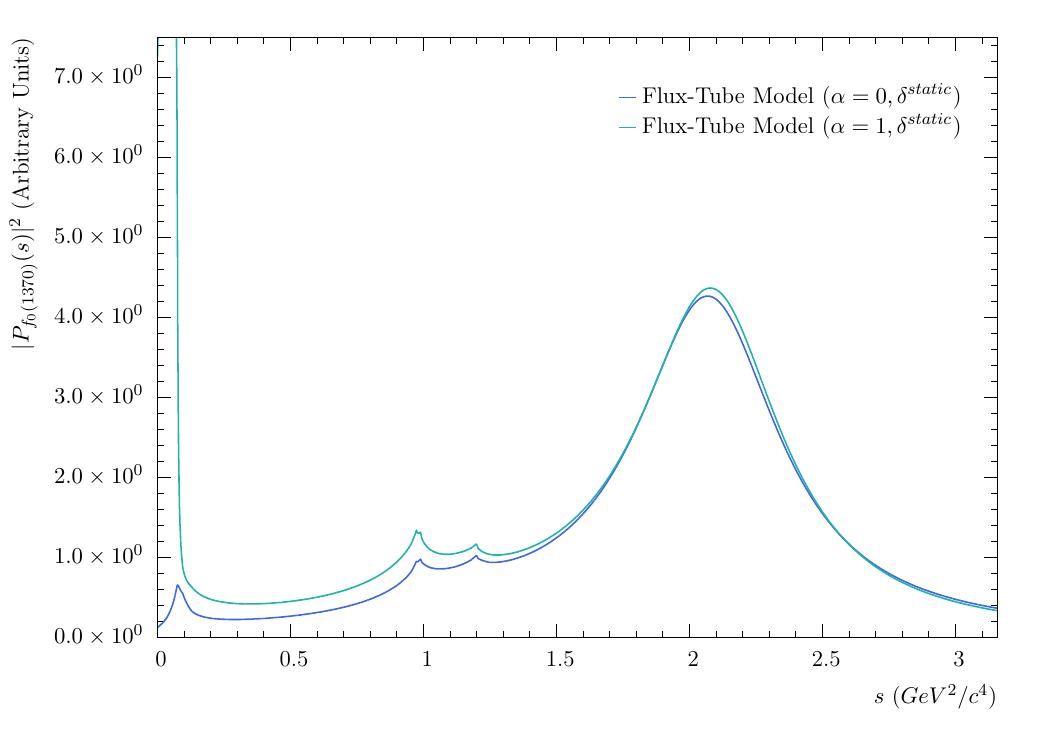}
  }
 \resizebox{500pt}{178pt}{
    \includegraphics{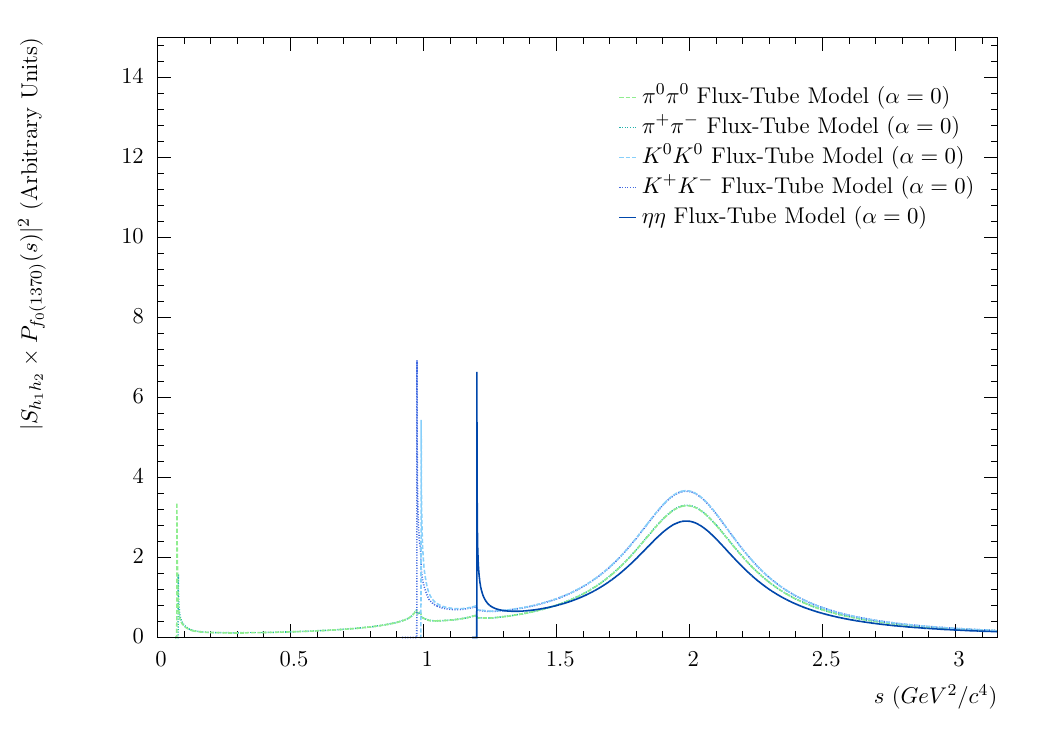}
    \includegraphics{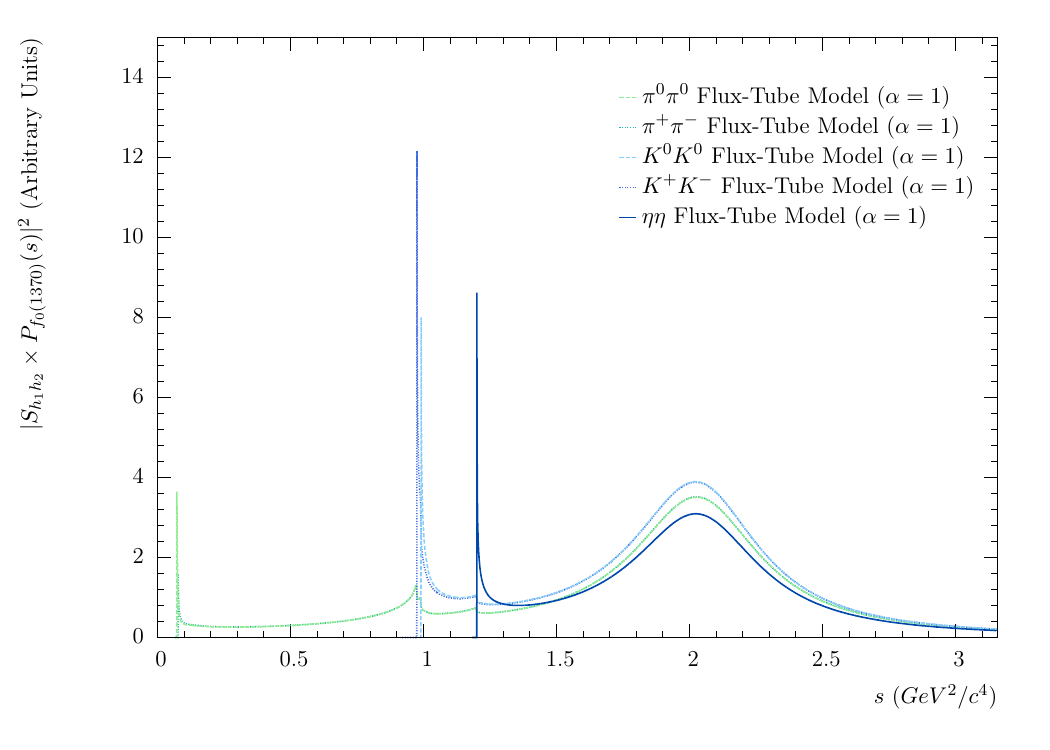}
  }
\end{center}
  \caption{A comparison of the $f_{0}(1370)$ propagator line shape with the $f_{0}(1370)\to\gamma\gamma$ channel included (top-left), with the $f_{0}(1370)\to\gamma\gamma$ channel included using the static
limit [$\delta^{static}$] (top-right) $f_{0}(1370)$ propagator with the wave-function
amplitude distortions for $\alpha=0$ (bottom-left) and $\alpha=1$ (bottom-right). The $f_{0}(1370)$ propagator is amplified through the dispersion relations near the $\pi^{0}\pi^{0}$
threshold where the only decay channel is $f_{0}(1370)\to\gamma\gamma$. For $\alpha=1$ this enhancement does not appear physical. Depending on the mixing in the strong annihilation process, the residual
QCD amplification from the wave-function amplitude distortion  may not  produce an amplification in the  $\eta\eta$ channel at threshold, but instead would contribute to the amplification of
 the $K^{+}K^{-}/K^{0}\bar{K}^{0}$ channels \cite{Weinstein:1990gu}.
 \label{fig:scalar}   }
\end{figure*}

\noindent The running of $\alpha_{s}(s)$ in the scalar amplitude $A({}_{1}^{3}P_{0}=0^{++})$ is taken into account using an effective phenomenological model of a saturated coupling in the low mass region 
\cite[Fig. 2]{Godfrey:1985xj}. The inclusion of the 
$f_{0}(1370)\to\gamma\gamma$ decay channel supresses the $\pi^{0}\pi^{0}$ and $\pi^{+}\pi^{-}$ threshold effects, as seen in Figure \ref{fig:scalar}. This can be seen in both the purely 
time-orders propagator and the propagator with non-resonant contributions,
\begin{equation}
\resizebox{0.375\textwidth}{!}{$
P(s)=\frac{\alpha}{s-\hat{m}^{2}(s)+\imath m_{0}\Gamma(s)}+\frac{1-\alpha}{2m_{0}\left(\sqrt{s}-\bar{m}(s)\right)+\imath m_{0}\Gamma(s)}
$}\label{eq:IMRpropogator}
\end{equation}
\noindent \cite{Nugent:2022ayu}.
The resulting $f_{0}(1370)\to\gamma\gamma$ decay width 
is $\Gamma(s)=2.4MeV$ which is comparable to the width extracted from Belle data Mushkelishvili-Omn\'es method (2.1keV)\cite{Pelaez:2015qba,Moussallam:2011zg,Mao:2009cc} and 
consistent with other predictions \cite[Fig. 40]{Pelaez:2015qba}.  For mesons composed of relativistic light quarks, the
contribution from the binding energy, $E_{B}=m-m_{Q}$, is non-negligible\footnote{$m_{Q}$ is the quark mass determined for a simple-harmonic oscillator wave-function within an approximate QCD potential 
which includes an exchange term + linear-confining and corrections for relativistic effects \cite{Godfrey:1985xj}.}. The static limit for the decay width \cite{Bergstrom:1982qv} is taken as an 
alternative to estimate the impact of the binding energy in the light-quark systems. From Figure \ref{fig:scalar}, it can be seen that in the static limit, the low mass region in the $f_{0}(1370)$ propagator 
is significantly enhanced. 

\section{Impact of the Low Energy QCD Potential}

\begin{figure*}[tbp]
\begin{center}
\resizebox{500pt}{178pt}{
    \includegraphics{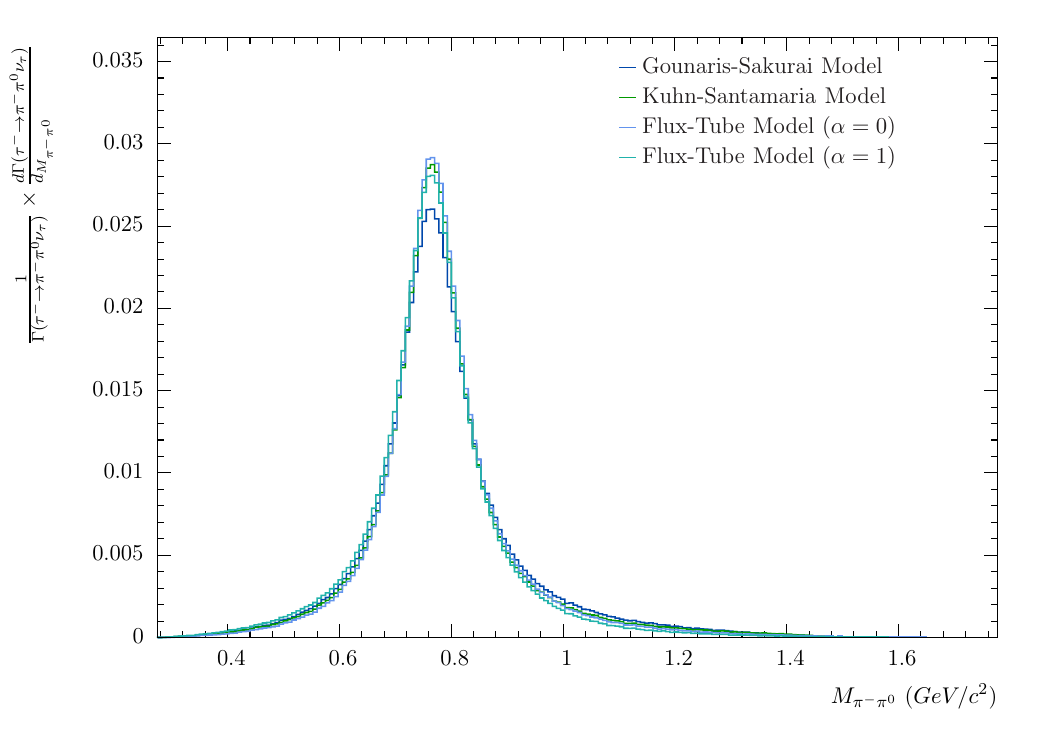}
    \includegraphics{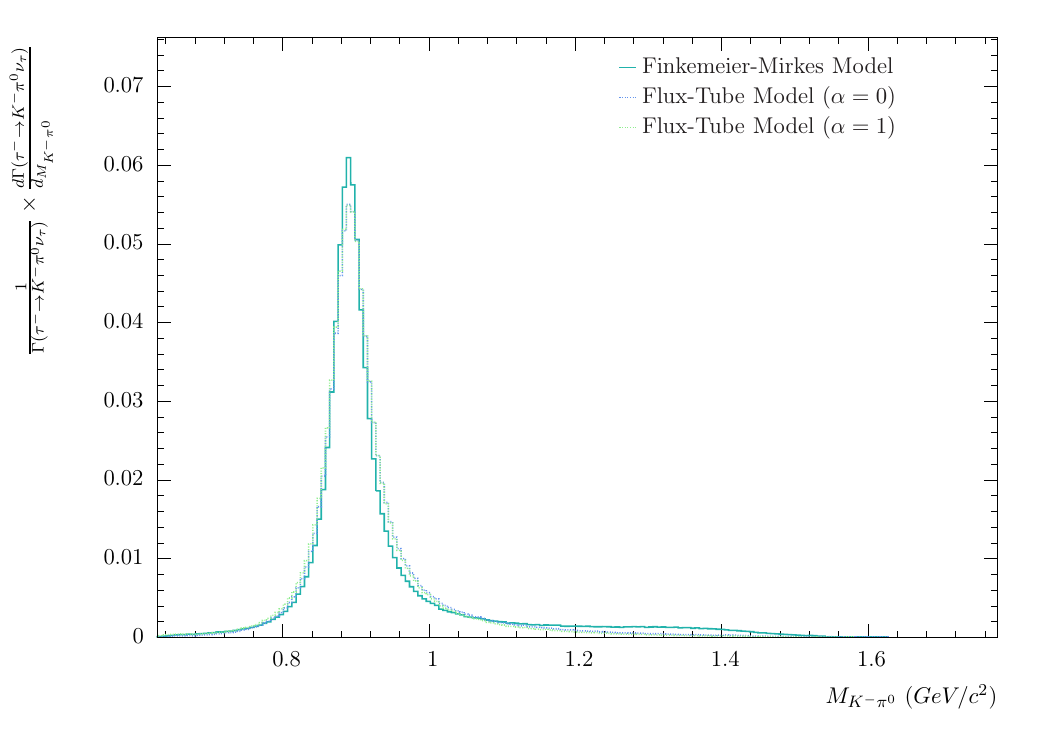}
  }
\resizebox{500pt}{178pt}{
    \includegraphics{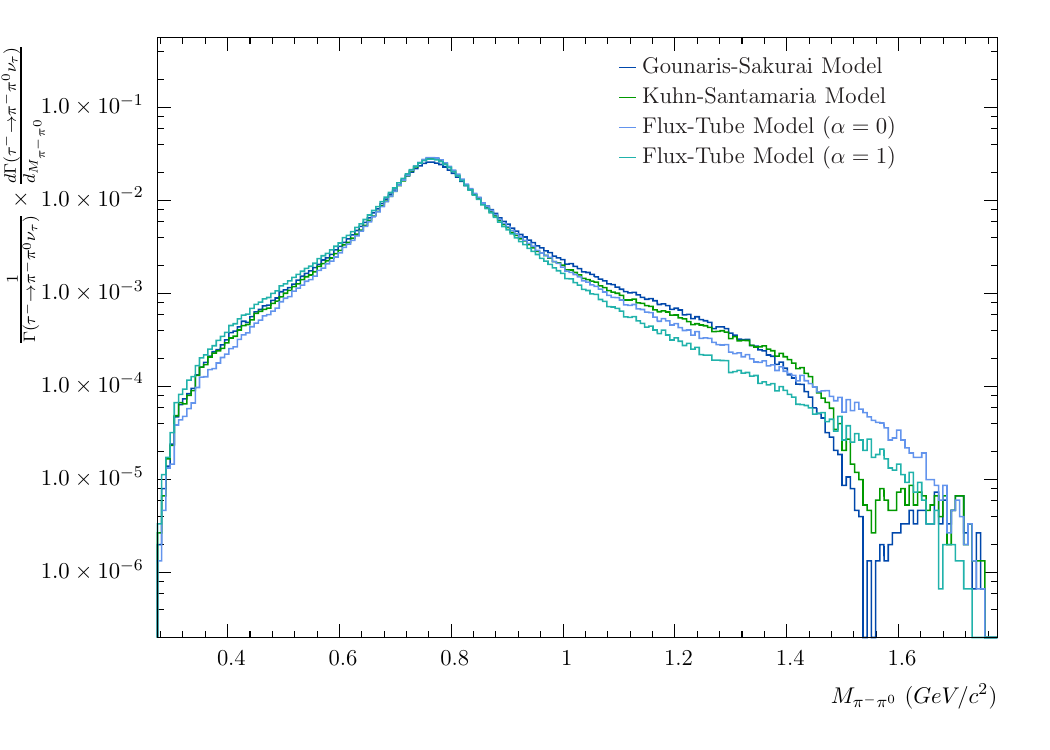}
    \includegraphics{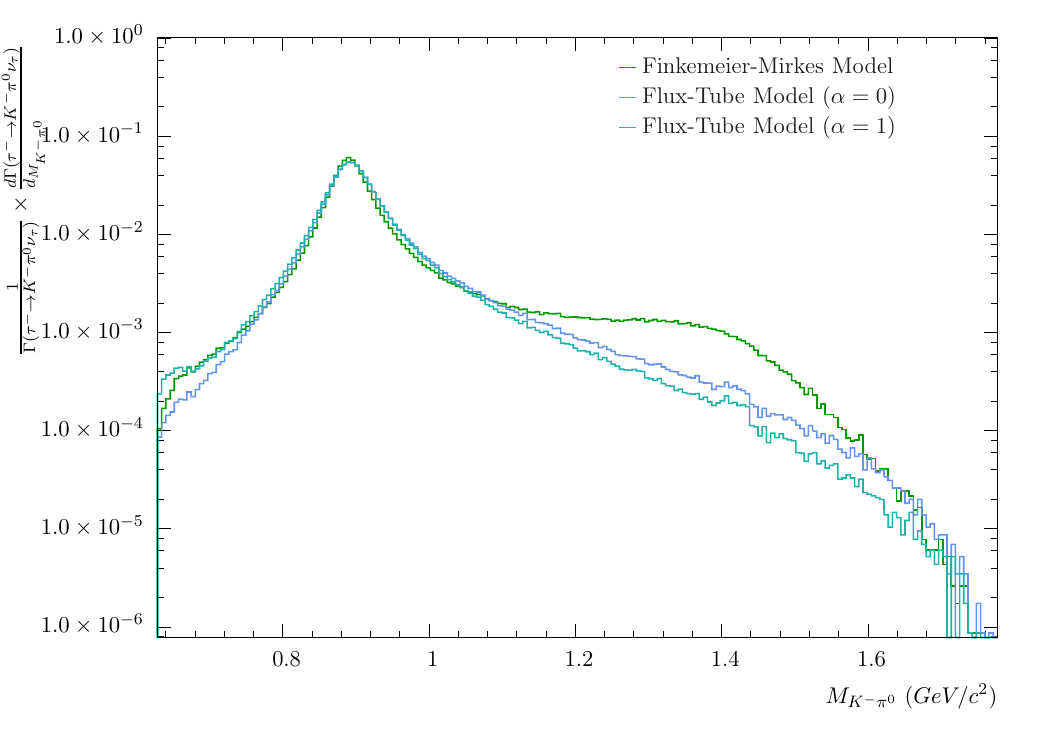}
 }
\end{center}
  \caption{The differential invariant mass spectra for $\tau^{-}\to\rho(770)\to\pi^{+}\pi^{0}\nu_{\tau}$ in linear (top-left) and log (botton-left) scale,
$\tau^{-}\to K^{*}(892)/K_{0}^{*}(1430)\to K^{+}\pi^{-}\nu_{\tau}$ in linear (top-right) and log (bottom-right) scale with the molecular wave-function distortion factor 
$S_{h_{1}h_{2}}^{2}$. In the $\tau^{-}\to\rho(770)\to\pi^{+}\pi^{0}\nu_{\tau}$, the  Gounaris-Sakurai Model \cite{Gounaris:1968mw} using the parameterization extracted 
from $e^{+}e^{-}\to\pi^{+}\pi^{-}\gamma$ data \cite{Lees:2012cj}, 
and the K{\"u}hn-Santamaria Model \cite{Kuhn:1990ad} with improved  parameterization \cite{Lees:2012cj} are super-imposed to illustrate the improvement in the agreement with other models, 
while the Finkemeier-Mirkes Model \cite{Finkemeier:1995sr} is super-imposed on the 
$\tau^{-}\to K^{*}(892)/K_{0}^{*}(1430)\to K^{+}\pi^{-}\nu_{\tau}$ channel.  The most significant difference between the Flux-Tube Breaking Model compared to the  Gounaris-Sakurai Model \cite{Gounaris:1968mw} 
and ChRL Models is due to the missing higher mass resonance. 
\label{fig:ScalarVectorSpectra} }
\end{figure*}

\begin{figure*}[tbp]
\begin{center}
\resizebox{500pt}{178pt}{
    \includegraphics{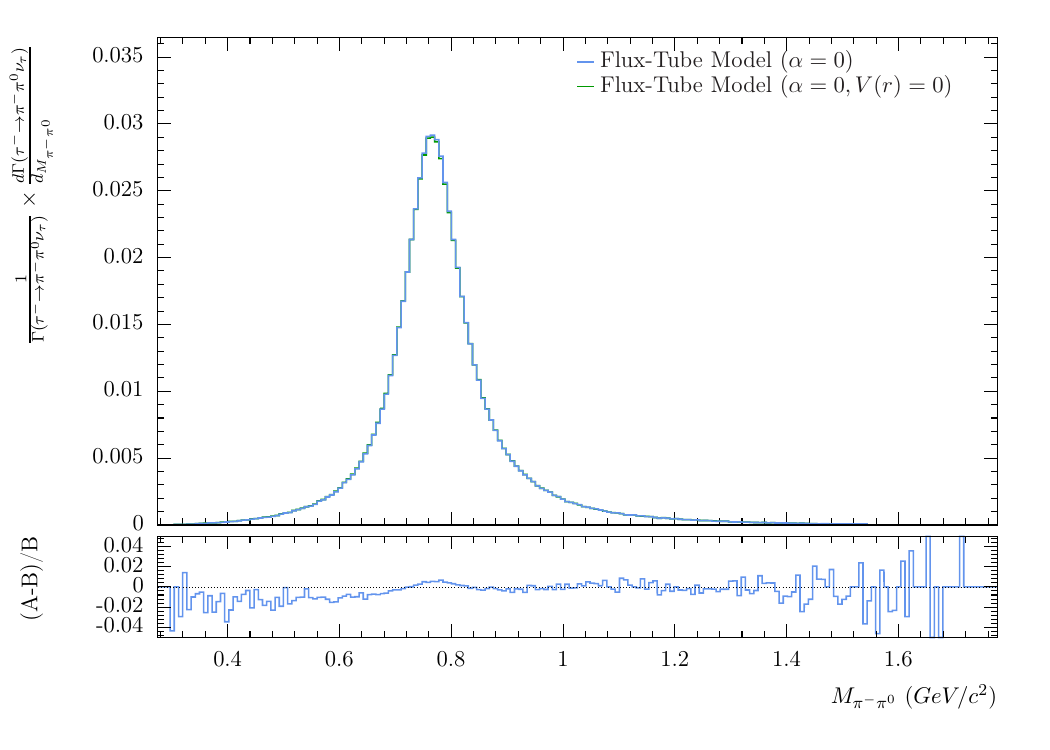}
    \includegraphics{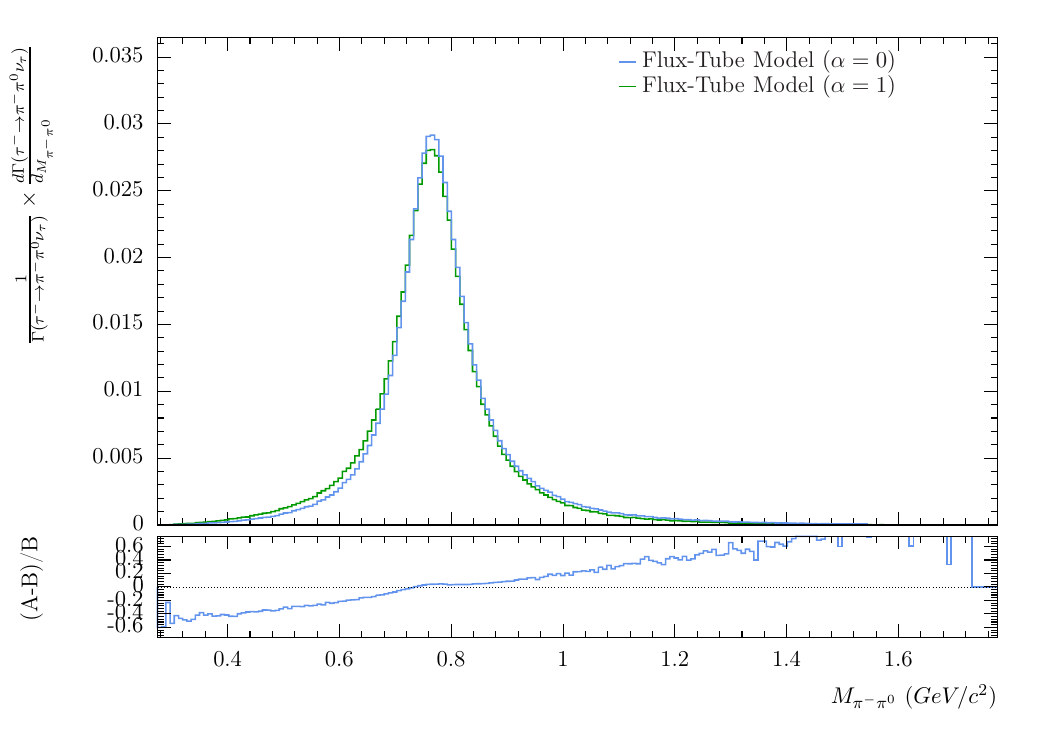}
  }
\end{center}
  \caption{A comparison of the differential invariant mass spectra for $\tau^{-}\to\rho(770)\to\pi^{+}\pi^{0}\nu_{\tau}$ in the Flux-Tube Breaking Model with and without the wave-function 
amplitude distortion (left) and with $\alpha=0$ and $\alpha=1$ (right) simulated for $1\times 10^{6}$ events. The lower plot represents the relative difference between the distributions. At BELLE-II,
the expected statistics after events selection for the $\tau^{-}\to\rho(770)\to\pi^{+}\pi^{0}\nu_{\tau}$ will be more than $100 \times$ greater than presented here \cite{abe2010belle,Kou_2019}.
\label{fig:rhoSpectraCompare} }
\end{figure*}

\begin{figure*}[tbp]
\begin{center}
\resizebox{333pt}{118pt}{
    \includegraphics{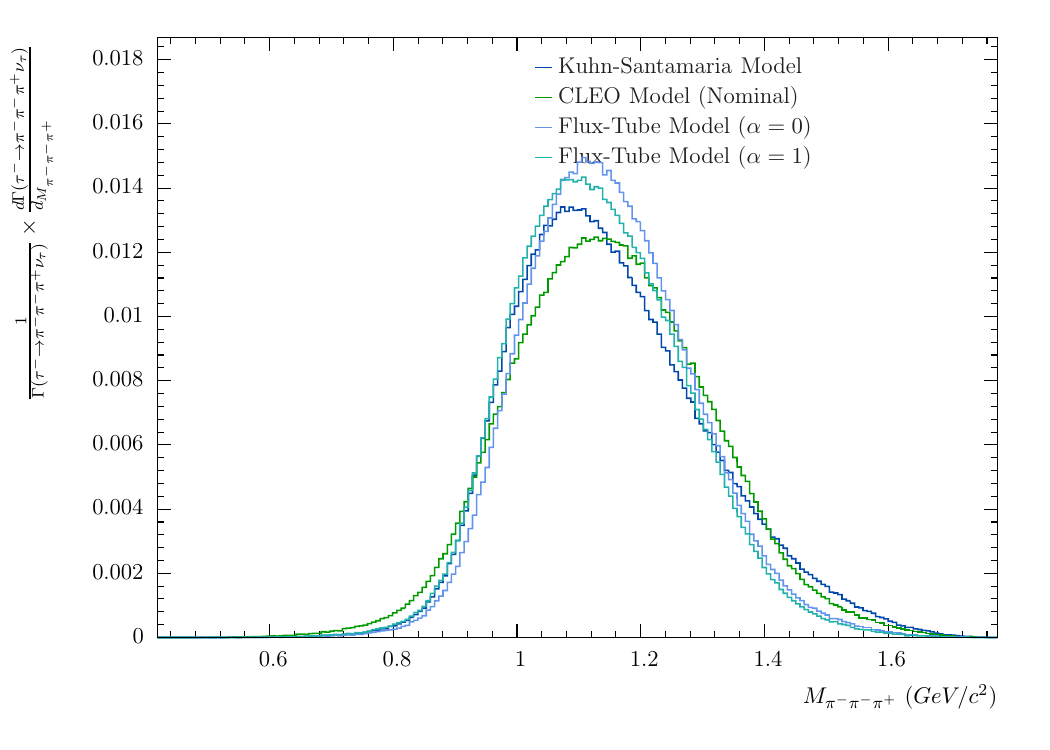}
    \includegraphics{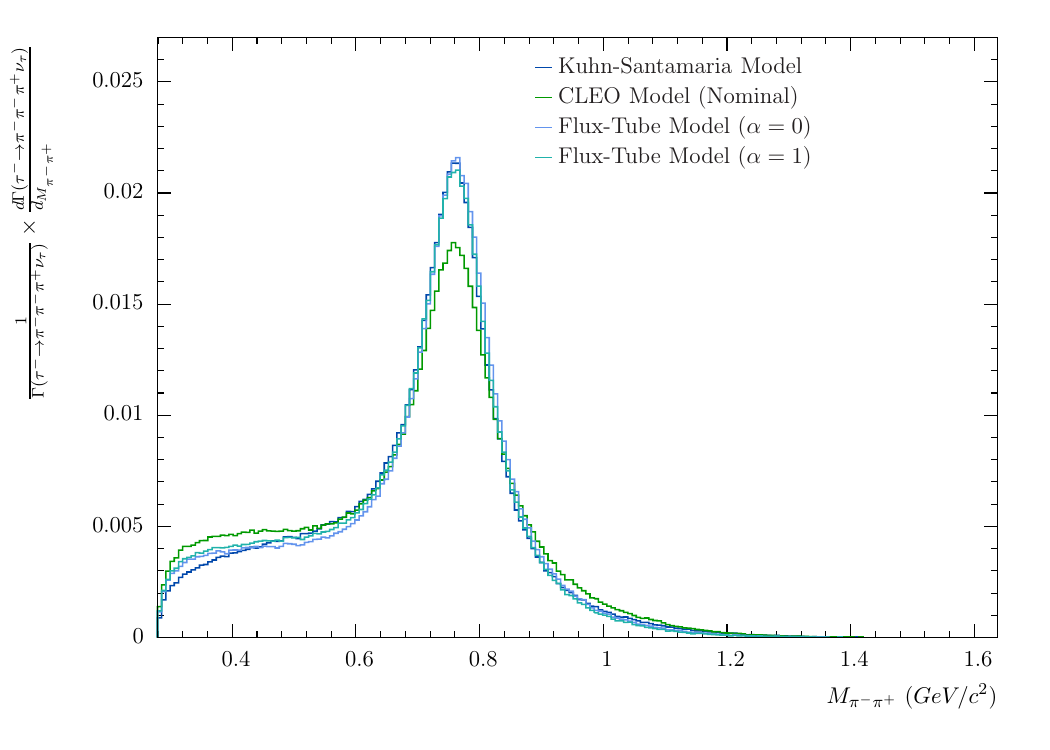}
  }
\resizebox{500pt}{118pt}{
    \includegraphics{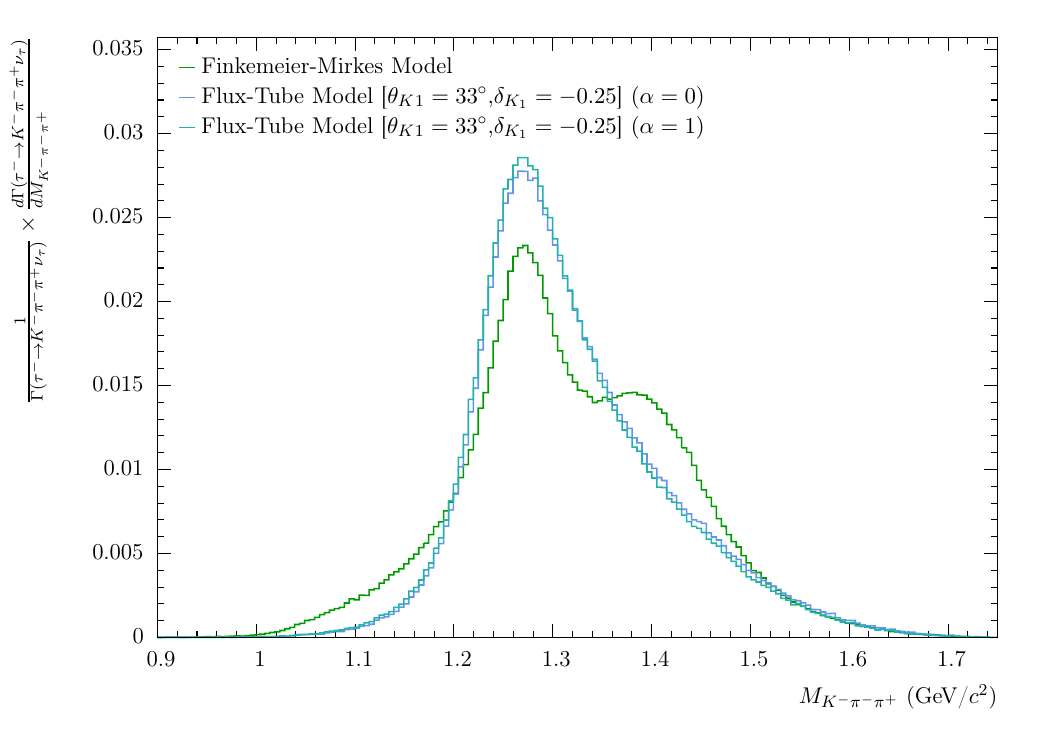}
    \includegraphics{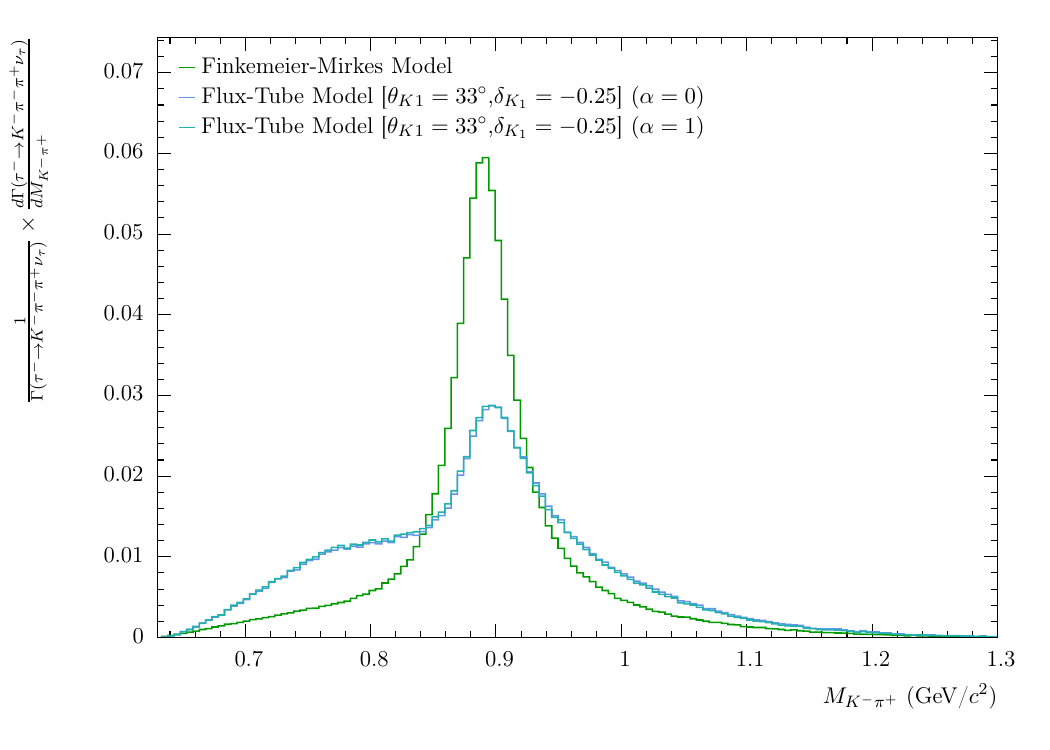}
    \includegraphics{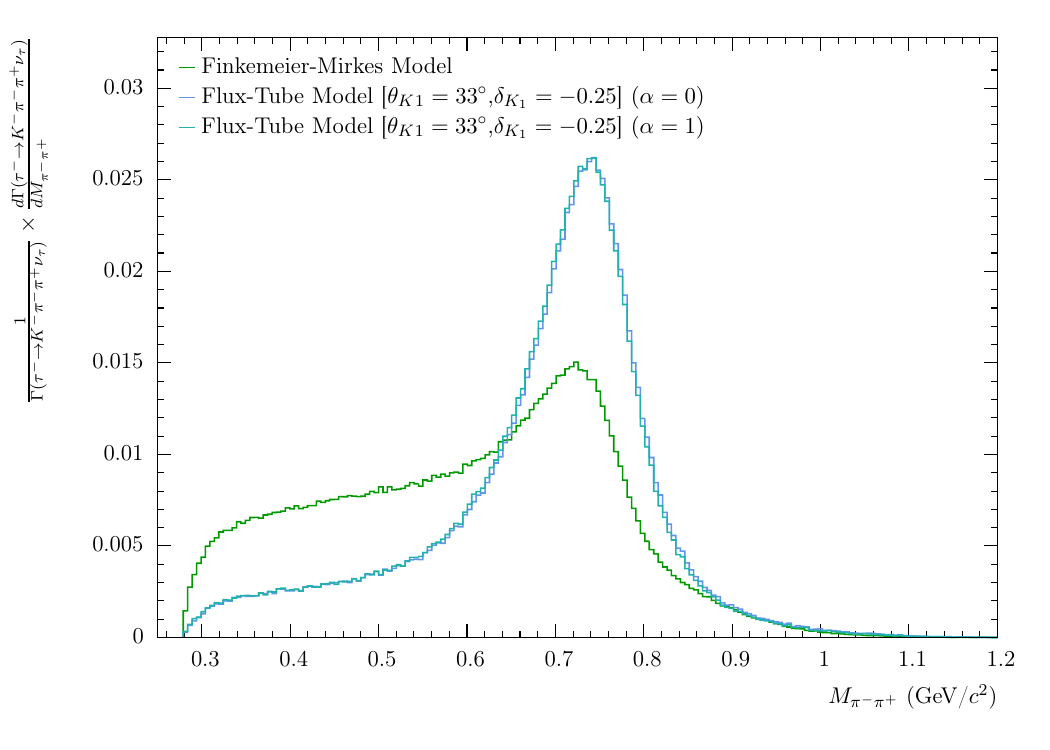}
  }
\end{center}
  \caption{The differential invariant mass spectra $\pi^{-}\pi^{-}\pi^{+}$ (top-left) and $\pi^{-}\pi^{+}$ (top-right) for the $\tau^{-}\to a_{1}(1260)\to\pi^{-}\pi^{-}\pi^{+}\nu_{\tau}$,
 and $K^{-}\pi^{-}\pi^{+}$ (bottom-left), $K^{-}\pi^{+}$ (bottom-middle) and  $\pi^{-}\pi^{+}$ (bottom-right)  $\tau^{-}\to K_{1}(1270/1400)\to K^{-}\pi^{-}\pi^{+}\nu_{\tau}$ 
with the molecular wave-function distortion factor $S_{h_{1}h_{2}}^{2}$. 
The  K{\"u}hn-Santamaria Model \cite{Kuhn:1990ad} and CLEO Model \cite{CLEO3pi} are  super-imposed to illustrate the improvement in the agreement of the Flux Tube Breaking Model
 when compared to the other models for
 the $\tau^{-}\to a_{1}(1260)\to\pi^{-}\pi^{-}\pi^{+}\nu_{\tau}$
channel. When compared to the experimental data \cite{Aubert:2007mh,Tau2012}, this improvement in the Flux-Tube Breaking Model is reflected through a tunable increase in the low mass $\pi^{+}\pi^{+}$
invariant mass spectra and narrower $a_{1}(1260)$ line-shape which is more consistent with the data \cite{Aubert:2007mh,Tau2012}. For the   $\tau^{-}\to K_{1}(1270/1400)\to K^{-}\pi^{-}\pi^{+}\nu_{\tau}$
channel, the Finkemeier-Mirkes Model \cite{Finkemeier:1995sr} is super-imposed on 
the $\tau^{-}\to K_{1}(1270/1400)\to K^{-}\pi^{-}\pi^{+}\nu_{\tau}$ channel. The disagreement in the $\pi^{+}\pi^{-}$ and
$K^{-}\pi^{+}$ spectra in the $K_{1}$ invariant mass distributions, when compared to \cite{Aubert:2007mh,Tau2012}, indicates that the value $\theta_{K_{1}}=33{}^{\circ}$ is inconsistent with the
experimental data.
 \label{fig:AxialVectorSpectra} }
\end{figure*}

Figures \ref{fig:ScalarVectorSpectra} and \ref{fig:AxialVectorSpectra} show a comparison of the simulated 2-hadron and 3-hadron decay spectra for the Flux-Tube Breaking Model 
with the wave-function amplitude distortions 
applied along with Chiral-Resonance-Lagrangian (ChRL) Models and Vector Dominance Models for comparison. The impact of the wave-function amplitude distortions on the $f_{0}(1370)$ propagator line-shape is illustrated 
in Figure \ref{fig:scalar} for each of the decay channels. Only the $\pi\pi$ channels 
show the double peak structure that was predicted in \cite{Nugent:2022ayu}. The large enhancement in the $KK$ channels explains why the measured $f_{0}(980)$ primarily decays through the $KK$ channels 
relative to the $\pi\pi$ channels. 
In \cite{Weinstein:1990gu}, it was shown that the $q\bar{q}q\bar{q}$ composition of the $f_{0}(980)$ and $a_{0}(980)$ may be formed through a mixture of meson-states
which is due to the amplitude of strong annihilation processes. In this case the
meson composition of the effective potentials must be constructed from a super-positioning of the meson potential. This would be manifested through the relative $KK$ and $\pi\pi$
branching fractions of the $f_{0}(1370)$ and $a_{0}(980)$ mesons, as observed in \cite{Barnes:1986uu,Barnes:1985cy,ODonnell:1985jss,Weinstein:1985ir}\footnote{The strong annihilation mixing
will also impact the $\eta\eta$ and $\eta^{\prime}\eta^{\prime}$ wave-function amplitude distortions, potentially shifting it below the  $\eta\eta$ and $\eta^{\prime}\eta^{\prime}$ productions threshold.}.
When interpreting the predictions, it is important to note which regions are relativistic and which are non-relativistic. The threshold regions which have the largest
 wave-function amplitude distortions are non-relativistic. When the particle becomes relativistic, the wave-function amplitude distortions tend to unity. Near the $\rho(770)$ peak and the $a_{1}(1260)$ peak
 for the $\rho(770)\pi$ channel, the outgoing mesons are relativistic. The $K^{*}(892)K$ contribution to the $a_{1}(1260)$ is non-relativistic. The outgoing mesons near the $K^{*}(892)$ and $\rho(770)$ 
peaks in the decays of the $K_{1}$ mesons are only quasi-non-relativistic with a $\beta\approx 0.5$. 
From Figure \ref{fig:ScalarVectorSpectra}, it can be seen that the P-wave $\tau^{-}\to\rho(700)\to\pi^{-}\pi^{0}\nu_{\tau}$ production for the Flux-Tube Breaking Model relative to 
the Gounaris-Sakurai Model \cite{Gounaris:1968mw,Lees:2012cj} in \cite{Nugent:2022ayu} are consistent depending on the fraction of purely time-ordered contribution to the propagator.
 From Figure \ref{fig:Enhancement}, the model dependence is sufficient 
for some discrimination between the model with the high statistics expected at Belle-II, however, this sensitivity may be limited by the knowledge of the fraction of purely 
time-ordered resonant contribution to the propagator, as seen in Figures \ref{fig:ScalarVectorSpectra} and \ref{fig:rhoSpectraCompare} near the  $\pi^{-}\pi^{0}$ threshold. Detailed studies are 
required to determine if the shape information can separate these two effects in individual decay processes.
  In Figure \ref{fig:AxialVectorSpectra}, for the $a_{1}(1260)$ and $K_{1}$ mesons, the wave-function amplitude distortion is primarily due to the colour hyper-fine spin-spin interaction 
in the $S_{1}$ production 
and therefore has a significant impact on the S/D-wave ratio near threshold of the vector and pseudo-scalar mesons, particularly for the $K_{1}$ mesons. This has to be
taken into account in any extraction of the mixing angle, $\theta_{K_{1}}$, and 
or the $SU(3)_{f}$ Flavour Breaking Factor $\delta_{K_{1}}$  \cite{Nugent:2022ayu,Suzuki:1993,Blundell:1995au,Asner:2000nx}.
In the $\pi^{+}\pi^{-}$ invariant mass, it can be seen that the wave-function amplitude distortions from the decay of the $a_{1}(1260)$ produced a significant impact on the 
$a_{1}(1260)$ $\pi^{-}\pi^{-}\pi^{+}$ invariant mass distribution. This mainly comes in through the modification of $\Gamma(s)$ in the propagator due to the wave-function amplification distortions, in
contrast to the $f_{0}(1370)$ where the wave-function amplitude distortions directly impact the cross-section.  The wave-function amplitude distortions from the decay of the $a_{1}(1260)$ also enhances the
low mass scalar region through enhancements of the $f_{0}(1370)$ and an improved $a_{1}(1260)$ line-shape, providing a plausible alternative explanation to the 
low mass scalar hypothesis proposed in \cite{Nugent:2013hxa}.
 
When extracting the distortion to the wave-function amplitude using the procedure in Section \ref{sec:ResidualQCD}, it was assumed that the wave-function has propagated to $r$ 
sufficiently large enough that it can be approximated by a plane wave. However, unlike the pseudo-scalar mesons, the vector meson, $\rho(770)$  and  $K^{*}(892)$ have a relatively short life-time. More 
specifically, the flight length of the $K^{*}(892)$ meson is $c\tau \beta=\frac{c \beta}{\Gamma}\approx 5fm\times \beta$, while for the $\rho(770)$ meson is
 $c\tau \beta=\frac{c \beta}{\Gamma}\approx 1.3fm\times\beta $.  These 
distances are comparable to the sizes of the outgoing mesons, $r_{\pi}=0.659\pm0.004fm$ and $r_{K}=0.56\pm0.031fm$ \cite{PDG2020}, and therefore the potentials. 
This means the wave-functions are still being modified by the inter-meson potentials when they decay and that the amplitude corrections do not fully describe the process.  
This is particularly important since, in Figure \ref{fig:AxialVectorSpectra}, it can be seen that the wave-function amplitude distortion in the S-wave channel at low energy in the decays of 
axial-vector mesons plays a significant role, in both the $a_{1}(1260)$ and $K_{1}(1270)$ line-shape. Moreover, from the decays of the $a_{1}(1260)$ and $K_{1}$ states, 
it can be seen that there is a non-negligible distortion to the $\rho(770)$ line shape when compared to the direct production mechanism,
 $\tau^{-}\to\rho^{-}(770)\nu_{\tau}$ and $\tau^{-}\to K^{*-}(892)\nu_{\tau}$.
Since the residual QCD interaction between mesons is not typically taken into account in the experimental measurements, it may explain why the $\rho(770)$ meson and
 $K^{*}(892)$ decay widths depend on the production mechanism \cite{PDG2020}. In the non-S-wave decay processes, it may be possible to probe the linear-confining potential and colour-Coulomb potential 
with multiple production mechanisms which would have different residual QCD potentials but with the same resonance shape.

\section{Conclusion} 

Residual QCD inter-meson potentials are found to have a non-negligible impact on the vector and axial-vector meson production in addition to the well known wave-function enhancements 
in the scalar states which can be interpreted as meson-meson molecular states \cite{Weinstein:1982gc,Weinstein:1983gd,Barnes:1986uu,Weinstein:1990gu,Isgur:1989js}. 
 We presented an improved model of the $f_{0}(1370)$ by including the missing $f_{0}(1370)\to\gamma\gamma$ decay channel. The addition of inter-meson residual QCD potentials due to the colour hyper-fine 
spin-spin interaction has a significant impact on the spectrum. Only the $\pi\pi$ decay channel would show the double peak structure results from the dispersion relations. This could be 
investigated in decays of the $a_{1}(1260)\to\pi^{-}f_{0}(1370)(\to\pi\pi)$ or in two-photon production $\gamma\gamma\to f_{0}(1370)(\to\pi\pi)$. The ratio of $\pi\pi$ and $KK$ production in 
$f_{0}(1370)$ near 0.980GeV ($f_{0}(980)$) is a signature for these inter-meson residual QCD potentials \cite{Barnes:1986uu}. There is already some supporting experimental evidence for 
the molecular description of the $f_{0}(500)$, $f_{0}(980)$ and $a_{0}(980)$ mesons \cite{Barnes:1986uu,Weinstein:1990gu,Isgur:1989js}.
 The exact enhancement of $\pi\pi$ and $KK$ production will depend 
on the quark composition predicted by the mixing of meson states, a consequence of the strong annihilation  \cite{Weinstein:1990gu}.
The residual inter-meson QCD potentials also have a non-negligible impact on the $a_{1}(1260)$ and $K_{1}$ state, 
particularly on the 3-body invariant mass distribution.  For the $a_{1}(1260)$, this represents a substantial improvement in the agreement between the Flux-Tube Breaking Model predictions 
in \cite{Nugent:2022ayu} and the data \cite{Aubert:2007mh}. 

\section*{Acknowledgement}
I would like to thank Zbigniew Was for drawing my attention to the discrepancy in the low mass region which is not well understood.
GCC Version 4.8.5 was used for compilation and the plots are generated using the external program GNUPlot \cite{gnuplot4.2}.

\footnotesize
\bibliography{paper}
\normalsize


\begin{table*}[b]
\begin{center}
\caption{The definition of the residual QCD potentials. The parameters are defined in Table \ref{table:Para}.
\label{table:Eq}}
\scriptsize
\begin{tabular}{p{4.5cm}p{7cm}}
\hline
Potential & Equation   \\
\hline
Spherical-Gaussian \cite{Barnes:1986uu} &
\begin{equation}
\resizebox{0.12\textwidth}{!}{$     
V(r)=V_{0}e^{-\frac{r^2}{2r_{0}^2}}  
$}
\label{eq:SG}
\end{equation}
\\
Shell Model (Woods-Saxon) \cite{walecka,Woods:1954zz}  & 
\begin{equation}
V(r)=V_{0}\left(\frac{1+e^{-\frac{a_{0}}{s_{0}}}}{1+e^{r-\frac{a_{0}}{s_{0}}}}\right)
\label{eq:WS}
\end{equation} \\
Shell Model (Parabolic) \cite{walecka}  &\begin{equation}
\resizebox{0.26\textwidth}{!}{$ 
V(r)=\begin{cases}                                                                                                                                                                                                0, & r>a_{0}\sqrt{2} \\
V_{0}\left(1-\left(\frac{r}{\sqrt{2}a_{0}}\right)^{2}\right),& \text{if}\ r<a_{0}\sqrt{2}
  \end{cases}
$}
\label{eq:PSM}
\end{equation}
 \\
Parabolic-Yukawa Model & 
\begin{equation}
\resizebox{0.27\textwidth}{!}{$
V(r)=\begin{cases}
V_{0}\left(\frac{e}{2}\right)  \frac{a_{0}}{r}e^{-\frac{r}{a_{0}}}, & r>a_{0} \\
V_{0}\left(\frac{1}{2}\right) \left(1+\left(1-\left(\frac{r}{a_{0}}\right)^{2}\right)\right), & \text{if}\ r<a_{0}
\end{cases} 
$}
\label{eq:PY}
\end{equation} \\
\hline
\end{tabular}
\end{center}
\end{table*}

\begin{table*}[b]
\begin{center}
\caption{The model parameters for the residual QCD potentials. The charge radius of the pion ($r_{\pi}$) and kaon ($r_{K}$) are from \cite{PDG2020}. The normalization of the strength of the 
residual QCD potentials, $V_{0}$, are based on \cite{Barnes:1986uu}. 
\label{table:Para}}
\scriptsize
\begin{tabular}{p{3.6cm}p{1.75cm}p{1.75cm}p{1.75cm}}
\hline
Decay & $V_{0}$ (GeV) &  $a_{0}$ & $s_{thickness}$  \\
\hline
$\pi^{0,\pm}$,$\rho^{0,\pm}$,$f_{0}(1370)$ & 0.4 & $r_{\pi}$ & $0.27$ \\
$K^{0,\pm}$,$K^{*(0,\pm)}(892)$,$K_{0}^{*}(1430)$ & 0.2 & $r_{K}$ & $0.27$ \\
$\eta$,$\eta^{\prime}(958)$ & 0.1 & $r_{K}/\sqrt(2)$ & $0.27$ \\
\hline
\end{tabular}
\end{center}
\end{table*}

\end{document}